\documentclass[preprint,aps,showpacs,nofootinbib]{revtex4}

\usepackage{epsfig,amssymb,amsmath}

\def\a{\alpha}

\newcommand{\beq}{\begin{equation}}
\newcommand{\eeq}{\end{equation}}
\newcommand{\bea}{\begin{eqnarray}}
\newcommand{\eea}{\end{eqnarray}}
\newcommand{\bear}{\begin{array}}
\newcommand {\eear}{\end{array}}
\newcommand{\bef}{\begin{figure}}
\newcommand {\eef}{\end{figure}}
\newcommand{\bec}{\begin{center}}
\newcommand {\eec}{\end{center}}

\newcommand{\la}{\left\langle}
\newcommand{\ra}{\right\rangle}

\def\REF#1{(\ref{#1})}
\def\GEV#1{10^{#1}{\rm\,GeV}}

\def\oten#1{ {\mathcal O}(10^{#1})}

\newcommand{\dnf}{\Delta N_{\rm eff}}

\begin{document}
\draft
\tighten
\preprint{TU-929}
\title{\large \bf
Axionic Co-genesis of Baryon, Dark Matter and Dark Radiation
}
\author{
    Kwang Sik Jeong\footnote{email: ksjeong@tuhep.phys.tohoku.ac.jp}
    and
    Fuminobu Takahashi\footnote{email: fumi@tuhep.phys.tohoku.ac.jp}}
\affiliation{
 Department of Physics, Tohoku University, Sendai 980-8578, Japan
    }

\vspace{2cm}

\begin{abstract}

We argue that coherent oscillations of the axion field excited by the misalignment mechanism
and non-thermal leptogenesis by the saxion decay can naturally explain the observed abundance
of dark matter and baryon asymmetry, thus providing a solution to the baryon-dark matter
coincidence problem.
The successful axionic co-genesis requires a supersymmetry breaking scale of $\oten{6-7}$\,GeV,
which is consistent with the recently discovered standard-model like Higgs boson of mass about
$126$\,GeV.
Although the saxion generically decays into a pair of axions, their abundance sensitively
depends on the saxion stabilization mechanism as well as couplings with the Higgs field.
We discuss various ways to make the saxion dominantly decay into the right-handed neutrinos rather
than into axions, and show that the abundance of axion dark radiation can be naturally as small as
$\dnf \lesssim {\cal O}(0.1)$, which is allowed by the Planck data.
\end{abstract}

\maketitle


\section{Introduction}

There are many coincidences in nature. Some of them may be just a coincidence or may be a consequence of
anthropic selection, while others may be due to a novel physical mechanism.
We consider the last possibility, since it will provide us with a hint of physics beyond the standard
model (SM).
In this paper we focus on the baryon-dark matter coincidence problem, i.e., why the dark matter density
is about five times larger than the baryon density,  $\Omega_{c} \sim 5 \Omega_b$. If these two
have a totally different origin, it is a puzzle why they happen to have the density in a similar size.

There have been many works on this long-standing puzzle.  Among them, the asymmetric dark matter scenario
has recently attracted great attention~\cite{Kaplan:2009ag}. If the dark matter particle has
asymmetry which is comparable to the baryon asymmetry,  the coincidence problem could be solved for the
dark matter mass of order the nucleon mass. Of course this is just a replacement of the problem.
The similarity of the baryon and dark matter abundances is replaced with the similarity of the nucleon mass
and the dark matter mass.  So, if the dark matter mass is independent of the QCD scale, the puzzle remains
unsolved. If the coincidence is due to a novel physical mechanism, therefore, the mechanism may be such that
the dark matter abundance  depends on the QCD scale, which determines the proton mass and thus
the baryon density.

The axion is a pseudo-Nambu-Goldstone boson associated with the spontaneous breakdown
of the Peccei-Quinn (PQ) symmetry, and it elegantly solves the strong CP problem when it couples
to the QCD anomaly~\cite{Peccei:1977hh,QCD-axion}.
The invisible axion is one of the plausible candidates for dark matter.
Interestingly, its mass and abundance depend on the QCD scale and the PQ breaking scale.
To our knowledge, the axion is the only dark matter candidate whose
abundance naturally depends on the QCD scale.\footnote{
Although its dependence on the QCD scale is not exactly same as the baryon density,
we will see that it may be related to another interesting coincidence between the QCD scale
and the light quark mass (or the weak scale).}
Therefore, we consider a possibility that the PQ sector is responsible for both the baryon asymmetry
and dark matter, i.e., {\it the axionic co-genesis}.

One particularly attractive mechanism that offers an explanation for the observed baryon
asymmetry is leptogenesis~\cite{Fukugita:1986hr,Buchmuller:2005eh}.
If the right-handed neutrinos are charged under the PQ symmetry, the axion as well as
its scalar partner, saxion, are naturally coupled to the right-handed neutrinos.\footnote{
In this case, the axion becomes a majoron~\cite{Langacker:1986rj}.
}
It is known that the saxion tends to dominate the energy density of the early universe.
Then, the saxion decays into right-handed neutrinos, whose CP-violating decays create
a lepton asymmetry~\cite{Lazarides:1991wu,Asaka:1999yd}. As we shall see shortly,
the saxion mass should be of $\oten{6-7}$\,GeV, for successful non-thermal leptogenesis.

The saxion generally decays also into a pair of axions.
Such non-thermally produced axions remain relativistic until present, giving contributions
to the effective number of neutrinos, $N_{\rm eff}$.\footnote{
The abundance of relativistic axions produced by flaton decays was studied in Ref.~\cite{Chun:2000jr}.
The late-time increase of $N_{\rm eff}$ by decaying particles (e.g. saxion to
two axions, and gravitino to axion and axino) was studied in
Ref.~\cite{Ichikawa:2007jv}. The string axion production from the modulus decay was
considered in Refs.~\cite{Higaki:2012ba,Higaki:2012ar,Cicoli:2012aq}.
See also Refs.~\cite{Fischler:2010xz,Hasenkamp:2011em,Menestrina:2011mz,Kobayashi:2011hp,Hooper:2011aj,Jeong:2012np,Choi:2012zna,
Graf:2012hb,Hasenkamp:2012ii,Bae:2013qr}
and many others.
The possibility that the $X$ particle was in thermal equilibrium was studied in
Ref.~\cite{Nakayama:2010vs}.
}
Some of the recent CMB observations gave a slight preference to the excess of
$N_{\rm eff}$, coined dark radiation~\cite{Hou:2012xq, Hinshaw:2012fq}.
On the other hand, the Planck results did not confirm such excess, and
place a constraint on $N_{\rm eff}$~\cite{Ade:2013lta},
\beq
N_{\rm eff} = 3.30^{+0.54}_{-0.51} ~~~(95\%;~{\rm Planck + WP + highL + BAO}).
\label{NeffPl}
\eeq
As we shall see shortly, although the axions are generically produced by the saxion decay,
its abundance sensitively depends on the saxion stabilization mechanism, and can be naturally
as small as $\Delta N_{\rm eff} ={\cal O}(0.1)$, which is still allowed by the Planck data.

Most of our results in this paper can be directly applied to non-supersymmetric cases, but we use
a supersymmetric (SUSY) language for simplicity. We shall return to the non-SUSY case
in the last section.  In fact, there is an interesting implication
in the SUSY framework.
In SUSY, the saxion acquires a mass $m_\sigma$ from
the SUSY breaking. For a generic K\"ahler potential, we expect $m_\sigma \sim m_{3/2}$ where
$m_{3/2}$ denotes the gravitino mass.
Interestingly,  high-scale SUSY breaking of $\oten{6-7}$\,GeV
is consistent with the recently discovered SM-like Higgs boson of mass about $126$\,GeV~\cite{:2012gk,:2012gu}
(see Refs.~\cite{Okada:1990gg,Giudice:2011cg}).
In fact, the observed Higgs boson mass as well as no experimental signatures of SUSY leads
to another puzzle: if SUSY is realized in nature, why its breaking scale is much higher
than the weak scale, which generically requires severe fine-tuning for the correct electroweak
symmetry breaking.
Our framework, the axionic co-genesis, may thus provide explanations for the puzzle, because
the non-thermal leptogenesis by the saxion decay would be impossible otherwise.

The rest of this paper is organized as follows. In Sec.~\ref{sec:2} we describe our set-up, and
then discuss the axionic co-genesis of baryon, dark matter and radiation in detail in Sec.~\ref{sec:3}.
The saxion stabilization is discussed in Sec.~\ref{sec:saxion-stabilization}.
The last section is devoted to discussion and conclusions.

\section{Set-up}
\label{sec:2}

The model we shall discuss is a PQ invariant extension of the supersymmetric SM
where PQ charged right-handed neutrinos ($N_i$ with $i=1,2,3$) are added to
implement the seesaw mechanism~\cite{seesaw}.
Thus $N_i$ becomes massive after the PQ symmetry breaking, and couples to the saxion
with a coupling proportional to its mass. Here and in what follows,
we assume that the lepton number is explicitly broken by adding couplings of the
PQ scalars with Higgs fields (such as (\ref{XHH})
or the second term of (\ref{see-saw})) in order to get rid of a massless majoron~\cite{Langacker:1986rj}.

We assume that there are multiple PQ scalar fields, $X_\alpha$, which
acquire a vacuum expectation value (VEV), $\langle X_\alpha \rangle = v_\alpha$.
The $X_\alpha$ can be expanded around its VEV as
\beq
X_\alpha \;=\; \left(v_\alpha +\frac{ \sigma_\alpha}{\sqrt{2}}\right)
\exp\left({\frac{ia_\alpha}{\sqrt{2} v_\a}}\right),
\eeq
where $\sigma_\alpha$ and $a_\alpha$ denote the radial and phase component of $X_\a$,
respectively. The axion $a$ and saxion $\sigma$ are expressed in terms of
those components of the PQ breaking fields,
\bea
a  &=& \sum_\alpha \frac{q_\alpha v_\alpha}{f} \,a_\alpha, \\
\label{saxion-field}
\sigma  &=& \sum_\alpha \frac{q_\alpha v_\alpha}{f} \,\sigma_\alpha,
\eea
where $q_\alpha$ is the PQ charge of $X_\alpha$, and the PQ scale $f$ is
determined by
\bea
f^2 = \sum_\alpha q^2_\alpha v^2_\alpha.
\eea

The interactions relevant to our discussion are written as
\bea
{\cal L} = \frac{x}{\sqrt2 f}\, \sigma (\partial^\mu a)\partial_\mu a
+ \left( \frac{1}{2} \lambda_i \sigma N_i N_i + \frac{1}{2}M_i N_i N_i
+ h_{ij} H_u N_i L_j + {\rm h.c.} \right),
\eea
below the PQ breaking scale $f$. Here  the $h_{ij}$ term generates Dirac neutrino masses
after electroweak symmetry breaking.
The saxion couples to the axions with a coupling~\cite{Chun:1995hc},
\bea
x = \sum_\alpha \frac{q^3_\alpha v^2_\alpha }{f^2}.
\eea
Here one should note that both $x$ and $f$ depend on the charge normalization, but
the combination $x/f$ does not.
In the following analysis, we assume a charge normalization such that all
$q_\alpha$ are order unity.

The right-handed neutrinos obtain masses from the PQ breaking, which
depend on how they couple to the PQ breaking fields.
For instance, one can take the PQ charge assignment that allows
the superpotential terms
\bea
\label{see-saw0}
W = \kappa X_\alpha N N + \kappa^\prime N L H_u,
\eea
where the family indices have been omitted.
Then $M_i\sim \kappa_i f$.
After integrating out the heavy right-handed neutrinos,  light neutrino
masses are generated as
\bea
m_\nu \sim \frac{\kappa^{\prime 2}}{\kappa}\frac{v^2}{f},
\eea
where $v$ is the Higgs vacuum expectation value.
We are interested in the case where at least one of the right-handed
neutrinos has a mass much smaller than the PQ breaking scale so that it can be
produced by saxion decay. As we shall see shortly,
the reference values are $f \sim \GEV{12}$ and the saxion mass $m_\sigma \sim \GEV{6}$.
This requires $\kappa$ to be smaller than $m_\sigma/f \sim 10^{-6}$,
which can be achieved if $N_i$ are charged under some flavor symmetry.
Another way to provide small masses to the right-handed neutrinos is to consider
higher-dimensional terms,\footnote{
One may instead consider the K\"ahler potential term,
$(X^\ast_\alpha/X_\beta) N N$, for the PQ fields having
$\langle X_\alpha \rangle \sim \langle X_\beta \rangle$.
Then the right-handed neutrino masses are induced after SUSY breaking
via the Giudice-Masiero mechanism \cite{Giudice:1988yz}, and naturally
similar to the saxion mass.
In this case, the saxion coupling for $\sigma N_iN_i$ is still proportional to
$M_i/f$ with a coefficient of order unity as long as $\alpha\neq \beta$.
}
\bea
\label{see-saw}
W = \kappa \frac{X^2_\alpha}{\Lambda}N N
+ \kappa^\prime \frac{X_\beta}{\Lambda} N L H_u,
\eea
by assigning appropriate PQ charges to the involved fields.
Here $X_\alpha$ can be in general different from $X_\beta$,
but we assume $\la X_\alpha\ra \sim \la X_\beta \ra$ for simplicity.
In this case, the right-handed neutrinos obtain masses about $\kappa f^2/\Lambda$,
leading to the light neutrino mass,
\bea
m_\nu \sim
\frac{\kappa^{\prime 2}}{\kappa}\frac{v^2}{\Lambda}.
\eea
The value of $m_\nu$ is determined by $v^2/\Lambda$, independent of $f$, in contrast to
the previous case.
The cut-off scale $\Lambda$ is considered to be around the GUT scale
(or the Planck scale if $\kappa \ll 1$), for which one can easily obtain the hierarchy $M_i\ll f$
as well as small neutrino masses consistent with the observational bounds.
Note that the axion coupling with left-handed leptons at low energy scales is suppressed
in this case if $\alpha=\beta$, in contrast to the previous case and the original Majoron model.

The saxion coupling to $N_i$ is proportional to $M_i$, and the numerical coefficient depends on the model.
To be concrete, we take the superpotential (\ref{see-saw}).
Then the relation between $\lambda_i$ and $M_i$ is fixed to be,\footnote{
In the model (\ref{see-saw0}),
$\lambda_i$ is two times smaller. Our main results hold also in this case without
significant modifications.
}
\bea
\lambda_i = \frac{\sqrt2 q M_i}{f},
\eea
where $q$ denotes the PQ charge of the PQ breaking field that is responsible
for the right-handed neutrino mass.
From the saxion couplings $x$ and $\lambda_i$, one finds that the branching ratio for
$\sigma\to N_iN_i$ is written
\bea
\label{Br-Ni}
{\rm Br}(\sigma\to N_i N_i)
= \frac{2}{r} \Big(\frac{2M_i}{m_\sigma}\Big)^2
\left(1-\Big(\frac{2M_i}{m_\sigma}\Big)^2 \right)^{3/2} B_a
<
\frac{0.36}{r} B_a,
\eea
if the process is kinematically allowed.
Here $m_\sigma$ is the saxion mass, and we have defined $r\equiv x^2/q^2$,
which is independent of the charge normalization.
$B_a$ denotes the branching ratio for the saxion decay into an axion pair,
\bea
B_a \simeq \frac{1}{\Gamma_\sigma}\frac{x^2}{64\pi}\frac{m^3_\sigma}{f^2},
\eea
with $\Gamma_\sigma$ being the total saxion decay rate.
The above shows that the upper bound on ${\rm Br}(\sigma\to N_i N_i)$ is
set by $B_a$ and $r$.
If $r$ is comparable to or smaller than $0.3$,
${\rm Br}(\sigma \to N_iN_i)\gtrsim B_a$ is allowed, and the saxion can dominantly
decay into right-handed neutrinos.
In Sec.~\ref{sec:saxion-stabilization}, we will show that such a value of $r$ is
obtained naturally in supersymmetric axion models.

Here and in what follows we do not consider the saxion decay into right-handed sneutrinos,
assuming that the decay is kinematically forbidden. Since the saxion mass arises from the SUSY breaking,
it can be comparable to the right-handed sneutrino mass. Even if the decay mode is allowed, the following
argument of non-thermal leptogenesis remains almost intact.\footnote{
The LSP abundance can be
affected. However, they will decay into the SM particles if the R-parity is violated.
}

Finally we mention the PQ mechanism solving the strong CP problem.
In order for this mechanism to work, the PQ symmetry should be anomalous under
the QCD so that the axion couples to gluons via
\bea
\label{agg}
{\cal L} = \frac{g^2_s}{32\pi^2}
\frac{a}{F_a} G^a_{\mu\nu}\tilde G^{a\,\mu\nu},
\eea
where the axion decay constant is given by $F_a=\sqrt2 f/N_{\rm DW}$.
The domain wall number $N_{\rm DW}$ is determined by the PQ anomaly coefficient,
and it counts the number of discrete degenerate vacua of the axion potential.
For the anomalous coupling, we need either to assign PQ charges to the Higgs
doublets or  introduce additional colored matter fields carrying PQ charges.
In the former case, an effective higgsino $\mu$ parameter is generated
after the PQ symmetry breaking, for instance, from the superpotential~\cite{Kim:1983dt}
\bea
W = \kappa_H \frac{X^2_\alpha}{\Lambda} H_u H_d.
\label{XHH}
\eea
It is important to note that the above superpotential induces saxion couplings to
higgsinos as well as the SM fermions through mixing with the Higgs, which can significantly
suppress $B_a$ depending on the values of $m_\sigma$ and the $\mu$ parameter.
This implies that $B_a+\sum_i {\rm Br}(\sigma\to N_i N_i)$ needs not to be equal to one
even for the case with $r$ of order unity.
Alternatively one may include color-charged matter fields $\Phi+\Phi^c$ which obtain
large masses from the coupling $X_\alpha \Phi\Phi^c$
in the superpotential.
Unless the coupling is suppressed, the saxion decay into $\Phi$ and $\Phi^c$
can be kinematically forbidden so that $B_a+\sum_i {\rm Br}(\sigma\to N_i N_i)$ becomes
close to unity.

\section{Cosmology}
\label{sec:3}

\subsection{Saxion dynamics}

Let us begin by discussing the cosmological role of the saxion. In the SUSY framework,
the saxion corresponds to a flat direction in the supersymmetric limit.
During inflation the potential receives Hubble-induced corrections, and therefore the minimum of
the potential is generally shifted away from the true vacuum.
After inflation the saxion starts to oscillate with a large amplitude when the Hubble parameter
becomes comparable to its mass. Furthermore, the saxion dynamics can be significantly affected
by thermal effects, and it is possible that the large saxion amplitude is
realized by thermal corrections through couplings with the heavy PQ quarks~\cite{Kawasaki:2011ym}.
Thus, the saxion tends to dominate the energy density of the universe unless the reheating of
the inflaton is extremely low.

In a class of the saxion stabilization models and also in the non-SUSY case, the initial oscillation
amplitude of the saxion is considered to be of order the PQ breaking scale $F_a$. The typical scale
of $F_a$ in our scenario is much smaller than the Planck scale, and so, it is non-trivial if the saxion
can dominate the universe before the decay.
Even in this case, if the saxion sits near the origin after inflation, thermal inflation may take place
and the saxion-dominant universe is realized.
The duration of the thermal inflation does not have to be very long, because we do not aim at solving
the moduli problem. Note however that in this case one needs to arrange $N_{\rm DW}=1$ to avoid
the domain wall problem, and that the axion decay constant is constrained to be smaller than
about $4\times 10^{10}$ GeV since otherwise the axions produced by unstable string-wall network
would overclose the universe~\cite{Hiramatsu:2010yu,Hiramatsu:2010yn}.

In the following we assume that the coherent oscillations of the saxion dominates the universe.
This sets the initial condition for the axionic co-genesis.

\subsection{Dark matter and dark radiation}

The axions are produced by the saxion decay and they constitute the dark radiation.
The axions thus produced are not thermalized because
the decoupling temperature is about $10^{9}{\rm GeV}\times (f/10^{11}\rm GeV)^2$~\cite{Graf:2012hb,Graf:2010tv},
which is many orders of magnitude larger than the saxion decay temperature
$T_\sigma$ for the parameter ranges studied here.
Thus the axions increase the effective number of neutrino species ($N_{\rm eff}$) by the amount
\cite{Jeong:2012np,Choi:1996vz}
\bea
\label{Neff}
\Delta N_{\rm eff} = \left.\frac{\rho_a}{\rho_\nu}\right|_{\,\nu\,{\rm decouple}}
= \frac{43}{7} \frac{B_a}{1-B_a}\left(
\frac{43/4}{g_\ast(T_\sigma)} \right)^{1/3},
\eea
where $g_\ast(T_\sigma)$ counts the relativistic degrees of freedom at the saxion decay.
In deriving the above expression, we have used the fact that it is the entropy in the comoving
volume that is
conserved when the light degrees of freedom changes at e.g. the QCD phase transition.
Thus,  for $B_a \simeq 0.1$, $0.2$, and $0.3$ and $g_\ast(T_\sigma) = 106.75$,
we obtain $\dnf \simeq 0.32$, $0.71$, and $1.2$, respectively.
As we shall see shortly, the value of $B_a$ naturally falls in the
range of ${\cal O}(0.1)$, and therefore the dark radiation with $\dnf = {\cal O}(0.1)$
is a robust prediction of this scenario.

\begin{figure}[t]
\begin{center}
\begin{minipage}{16.4cm}
\centerline{
{\hspace*{0cm}\epsfig{figure=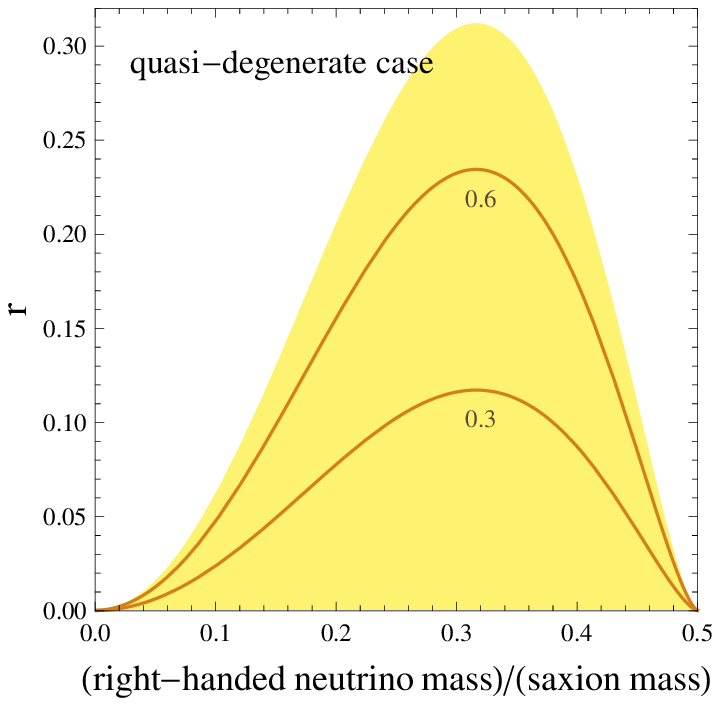,angle=0,width=7.4cm}}
{\hspace*{.4cm}\epsfig{figure=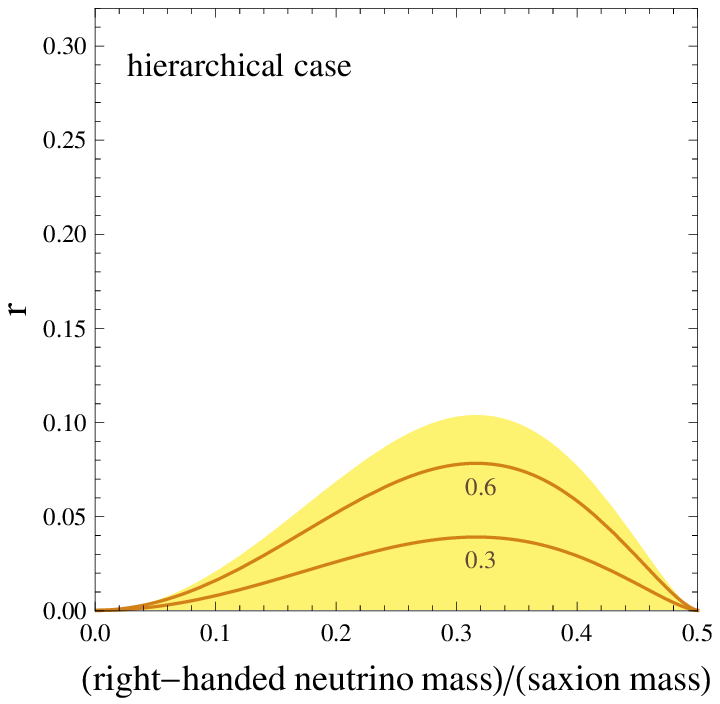,angle=0,width=7.4cm}}
}
\caption{
Axion dark radiation $\Delta N_{\rm eff}$ in the case with quasi-degenerate (left panel)
and hierarchical (right panel) right-handed neutrinos.
The amount of axion dark radiation is fixed by $B_a$, on which we have imposed
the condition $B_a+{\rm Br}(\sigma\to gg)+\sum_i {\rm Br}(\sigma\to N_iN_i)\simeq 1$
under the assumption that other saxion decay modes have small branching ratios.
The shaded region corresponds to $\Delta N_{\rm eff}\leq 0.8$, and the brown lines
represent the contours for $\Delta N_{\rm eff}=0.3$ and 0.6, from below, respectively.
For the hierarchical $N_i$ case, we have taken $M_1<m_\sigma/2\ll M_2,M_3$.
}
\label{fig:dark-radiation}
\end{minipage}
\end{center}
\end{figure}

Fig. \ref{fig:dark-radiation} shows how $\Delta N_{\rm eff}$ depends on $r$ and the mass of
the right-handed neutrino.
Here we have assumed that the saxion decays dominantly into right-handed
neutrinos, axions, and/or gluons, i.e.,
\bea
B_a + {\rm Br}(\sigma \to gg) + \sum_i {\rm Br}(\sigma\to N_iN_i)\simeq 1.
\eea
The process $\sigma\to gg$ is mediated by the saxion coupling to gluons,
which is analogous to the anomalous axion coupling to gluons \REF{agg}.
Thus ${\rm Br}(\sigma\to gg)$ is negligibly small compared to $B_a$ as long as
$r \gg \alpha^2_s/\pi^2 = {\cal O}(10^{-3})$.
In the yellow shaded region, we have $\Delta N_{\rm eff}\leq 0.8$, which
is the case when $B_a$ is smaller than about 0.22: the left panel is obtained
for quasi-degenerate right-handed neutrinos with
$M_1 \sim M_2 \sim M_3 < m_\sigma/2$, while the right panel is for
the hierarchical mass spectrum of $N_i$ with $M_1<m_\sigma/2 \ll M_2, M_3$.
The brown lines are the contours of $\Delta N_{\rm eff}=0.3$ and $0.6$, respectively.

On the other hand, the axion obtains a mass through QCD non-perturbative effects,
and starts to oscillate coherently when the Hubble parameter becomes comparable
to its mass.
These coherent axions constitute the cold dark matter of the universe with the
relic energy density given by \cite{Turner:1985si}
\bea
\Omega_a h^2 \simeq 0.18\, \theta^2_a \left(
\frac{F_a}{10^{12}{\rm GeV}} \right)^{1.19}
\left(\frac{\Lambda_{\rm QCD}}{400{\rm MeV}}\right)^{3/4},
\eea
where $\theta_a$ is the initial misalignment angle, and $\Lambda_{\rm QCD}$ denotes
the QCD scale. Note that we extracted the dependence of the QCD scale not only from
the thermal effects on the axion mass \cite{Gross:1980br} but also from
the zero-temperature axion mass.
Thus, the coherent axions produced by an initial misalignment $\theta_a ={\cal O}(1)$
make up the major part of cold dark matter in the universe for the axion decay constant
around $5\times 10^{11}$ GeV.
For a larger (smaller) value of $F_a$, we need to fine-tune the initial position close
to the potential minimum (maximum).

\subsection{Non-thermal leptogenesis}

A natural way to generate a baryon asymmetry is leptogenesis via the decays
of right-handed neutrinos.
To produce a sufficient baryon asymmetry in the saxion-dominated universe,
we consider non-thermal leptogenesis.
To proceed, let us estimate the saxion decay temperature.
This is determined by the saxion decay rate into particles that thermalize
to form a thermal bath:
\bea
T_\sigma \simeq \left(\frac{90}{\pi^2g_\ast(T_\sigma)}\right)^{1/4}
\sqrt{\Gamma_\sigma M_{Pl}},
\eea
from which it follows,
\bea
\frac{T_\sigma}{m_\sigma} \simeq 0.13|x|
\left(\frac{0.2}{B_a}\right)^{1/2}
\left(\frac{106.75}{g_\ast(T_\sigma)}\right)^{1/4}
\left(\frac{m_\sigma}{10^6{\rm GeV}}\right)^{1/2}
\left(\frac{10^{12}{\rm GeV}}{f}\right).
\eea
The above relation holds if the decay rate of $N_i$ into leptons and Higgs is much
larger than $\Gamma_\sigma$ so that the right-handed neutrinos decay almost instantly
after production.
This is indeed the case when $T_\sigma$ is smaller than $m_\sigma$, and
the branching ratio for $\sigma\to N_iN_i$ is sizable.
The decay rate of $N_i$ into leptons is larger than $\Gamma_{\sigma\to N_iN_i}$ by a factor
of approximately $(f/v)^2 m_\nu/m_\sigma$, which is about
$10^2\times r\,(m_\sigma/T_\sigma)^2$ for $B_a$ around 0.2 and $m_\nu\sim 0.1$ eV.
Thus in the parameter space with $r$ around 0.1 and $T_\sigma\lesssim m_\sigma/2$,
which is of our interest, the relation for $T_\sigma$ remains valid.\footnote{
More precisely, $\Gamma_\sigma$ should be replaced by $(1-B_a)\Gamma_\sigma$
since the axions produced in saxion decays are not thermalized as we will see shortly.
However this changes the result only slightly for small $B_a$.
}

The right-handed neutrinos are non-thermally produced by saxion decays when
$T_\sigma$ is low enough:
\bea
\label{non-thermal-leptogenesis}
T_\sigma \lesssim M_i < \frac{m_\sigma}{2},
\eea
thereby requiring $T_\sigma$ smaller than $m_\sigma/2$.
The lepton asymmetry generated by the decays of $N_i$ is partially converted
into baryon asymmetry through the sphaleron process according to $n_B=c_{\rm sph}n_L$,
where $c_{\rm sph}=-28/79$ for $T_\sigma\gg 10^2$ GeV in the SM.\footnote{
The saxion mass is considered to be of order the gravitino mass.
If all the SUSY particles as well as heavy Higgs bosons have similar masses,
there are only SM particles in the plasma at the electroweak phase transition.}
The present baryon asymmetry is then obtained to be
\bea
\frac{n_B}{s} =
c_{\rm sph} \frac{3}{2}\frac{T_\sigma}{m_\sigma}
\sum_i \epsilon_i {\rm Br}(\sigma\to N_iN_i),
\eea
for $T_\sigma\lesssim M_1$, where $s$ is the entropy density.
The condition $T_\sigma\lesssim M_1$ is necessary to avoid strong wash-out of
the lepton asymmetry produced by $N_i$ decays.
The CP asymmetry $\epsilon_i$ in the $N_i$ decay receives one-loop vertex and self-energy
contributions \cite{Covi:1996wh}, and crucially depends on the right-handed neutrino masses
and thus on $f$.

Let us estimate the baryon asymmetry generated by non-thermal leptogenesis.
There are two cases of interest, depending on the mass spectrum of $N_i$.
First, suppose hierarchical $N_i$.
Then an adequate baryon asymmetry is obtained when $N_1$ has mass not much smaller than
$m_\sigma/2$ while the other two have masses larger than $m_\sigma/2$.
This is because $\lambda_i$ is proportional to $M_i$, and $T_\sigma$ less than
$M_1$ is needed to avoid strong washout effects.
For $M_1<m_\sigma/2\ll M_2,M_3$, the CP asymmetry induced by the $N_1$ decay is given by
\bea
\epsilon_1  = \frac{3}{16\pi}\frac{m_{\nu_3}M_1}{v^2}\, \delta_{\rm eff},
\eea
where $|\delta_{\rm eff}|\leq 1$ is the effective leptogenesis CP phase,
and $m_{\nu_3}$ is the heaviest neutrino mass.
Taking $\delta_{\rm eff}=1$, the baryon asymmetry reads
\bea
\label{hierarchical-N}
\frac{n_B}{s} \simeq
10^{-11} \left(\frac{T_\sigma/m_\sigma}{0.2}\right)
\left(\frac{m_{\nu_3}}{0.05{\rm eV}}\right)
\left(\frac{M_1}{10^6{\rm GeV}}\right)
\left(\frac{{\rm Br}(\sigma\to N_1N_1)}{0.8}\right),
\eea
where an enhancement of a factor 2 can be obtained if the saxion is kinematically
allowed to decay into the right-handed sneutrinos.
On the other hand, it is also possible to have $N_i$ with masses close to each other.
For such quasi-degenerate $N_i$, the induced CP asymmetry
is approximately given by
\bea
\epsilon_i \approx
\frac{1}{16\pi}\frac{m_{\nu_3}M}{v^2} \frac{M}{\Delta M}\,\delta_{\rm eff},
\eea
where the right-handed neutrinos have masses around $M$ with a small splitting $\Delta M$.
The CP asymmetry is enhanced by a factor $M/\Delta M$, compared to the case with
hierarchical $N_i$.
As a result, the baryon asymmetry produced in the decays of three right-handed neutrinos
is approximately given by
\bea
\label{degenerate-N}
\frac{n_B}{s} \approx
10^{-10} \left(\frac{T_\sigma/m_\sigma}{0.2}\right)
\left(\frac{m_{\nu_3}}{0.05{\rm eV}}\right)
\left(\frac{M}{10^6{\rm GeV}}\right)
\left(\frac{M/\Delta M}{10}\right)
\left(\frac{B_N}{0.8}\right),
\eea
for $\delta_{\rm eff}$ close to 1, where $B_N \equiv \sum_i {\rm Br}(\sigma\to N_iN_i)$.

The induced baryon asymmetry depends on $m_\sigma$, $M_i$, $f$
and $B_a$, some of which also determine the amounts of axion dark matter and
dark radiation.
In the next subsection, taking into account such relations, we will explore
the region of parameter space where the baryon asymmetry observed in the universe
is explained by (\ref{hierarchical-N}) or (\ref{degenerate-N}).

\subsection{Axionic Co-genesis}

To get a sufficient baryon asymmetry, the saxion needs to decay mainly into a pair
of right-handed neutrinos.
The relation (\ref{Br-Ni}) tells that this is achieved when $B_a$ is not small,
under the natural assumption that $r$ is not much smaller than 0.1.
This implies that $\Delta N_{\rm eff}$ due to axion dark radiation generally
lies in the range between about 0.1 and 1 in our scenario,
which is consistent with the recent Planck results (\ref{NeffPl}).

Let us move on to the baryon asymmetry and dark matter.
It is important to note that an upper bound on $m_\sigma$ comes from the requirement
that $T_\sigma$ should be smaller than $M_1$ for given $f$,
while a lower bound arises from the requirement of generating adequate baryon asymmetry,
provided that the right-handed neutrinos are not extremely degenerate in mass.
The coherent axions produced by an initial misalignment $\theta_a ={\cal O}(1)$ make
up the major part of cold dark matter in the universe for the axion decay constant
around $5\times 10^{11}$ GeV.
For the PQ breaking scale in that range, $T_\sigma$ will be low enough to avoid strong
wash-out if the saxion mass is smaller than about $10^7$ GeV.
On the other hand, from (\ref{hierarchical-N}) and (\ref{degenerate-N}), we see
that $n_B/s\sim 10^{-10}$ is naturally obtained when the right-handed neutrino,
whose decay generates the lepton asymmetry, has a mass around or larger than
$10^6$ GeV.
Therefore the baryon asymmetry and dark matter of the present universe
can be achieved simultaneously for
\bea
F_a \sim 5\times 10^{11}\,{\rm GeV} &\,{\rm and}\,&
10^6\,{\rm GeV} \lesssim m_\sigma \lesssim 10^7\,{\rm GeV},
\eea
thereby providing a natural explanation to the baryon-dark matter coincidence
problem.
This also indicates high scale SUSY breaking around $10^{6-7}$ GeV because
the saxion is massless in the supersymmetric limit.
Interestingly, this is the SUSY breaking scale needed to accommodate a 126 GeV
Higgs boson within the minimal supersymmetric SM without large stop mixing.

\begin{figure}[t]
\begin{center}
\begin{minipage}{16.4cm}
\centerline{
{\hspace*{0cm}\epsfig{figure=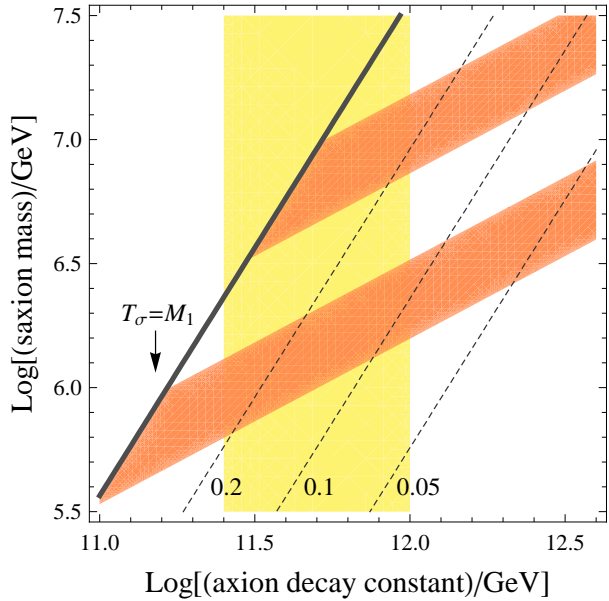,angle=0,width=8cm}}
}
\caption{
Axionic co-genesis in the saxion-dominated universe.
The upper (lower) orange region is the region where $n_B/s=(1\pm0.5)\times 10^{-10}$
is obtained via non-thermal leptogenesis with hierarchical (quasi-degenerate)
right-handed neutrinos.
Here we have taken $B_a=0.1$, for which the axion dark radiation gives $\Delta N_{\rm eff}\simeq 0.32$,
and have assumed that the saxion decays into right-handed neutrinos
with $B_N=0.9$ (see the text).
The relic axion density due to vacuum misalignment is consistent
with the observed cold dark matter abundance in the yellow region
for an initial misalignment angle of order unity.
In the region above the gray thick line, $T_\sigma$ is larger than $M_1$, and
the lepton asymmetry produced by $N_i$ decays is strongly washed out.
The dashed line represents $T_\sigma/m_\sigma=0.2,\,0.1,\,0.05$, from left,
respectively.
}
\label{fig:cogenesis}
\end{minipage}
\end{center}
\end{figure}

Fig. \ref{fig:cogenesis} illustrates how non-thermal leptogenesis works in the
saxion-dominated universe.
Here we have taken the model parameters as follows, for the cases with hierarchical
and quasi-degenerate right-handed neutrinos:
\bea
\mbox{\it hierarchical $N_i$} &:&
M_1 = 0.4 m_\sigma \ll M_2, \quad {\rm Br}(\sigma\to N_1 N_1)= 0.9,
\nonumber \\
\mbox{\it quasi-degenerate $N_i$} &:&
M = 0.4 m_\sigma,
\quad M/\Delta M=10,
\quad B_N = 0.9,
\eea
and fixed $B_a=0.1$ for both cases, under the assumption that $r$ lies in the range
between about 0.1 and 0.2.
In addition, we have taken $x=0.35 N_{\rm DW}$ for the PQ charge normalization such that
the PQ charges of $X_\alpha$ are co-prime to each other, for which $r$ is given by
$r\simeq 0.12 (N_{\rm DW}/q)^2$.
For hierarchical (quasi-degenerate) $N_i$ with the above properties,
$n_B/s=(1\pm 0.5)\times 10^{-10}$ is obtained in the upper (lower) orange region.
The observed baryon asymmetry can thus be explained within this region by slightly
changing the properties of $N_i$ or $\delta_{\rm eff}$.
In the region above the gray thick line, the saxion decay temperature is higher than
the right-handed neutrino mass, and the situation is reduced to the thermal leptogenesis
scenario.
The successful thermal leptogenesis will require the saxion to have mass of order
$10^9$ GeV for $F_a\sim 10^{11}$ GeV.
On the other hand, in the yellow region, the coherent axions produced around
the QCD phase transition can naturally account for the cold dark matter in the universe.
Finally we note that the axion dark radiation is also produced in saxion decays, for instance,
$\Delta N_{\rm eff}\simeq 0.32$ for $B_a=0.1$.
The amount of baryon asymmetry does not change much for $0.1\lesssim B_a \lesssim 0.2$
as long as the saxion dominantly decays into right-handed neutrinos.
From these we conclude that the axonic co-genesis is realized in the region where
the orange and yellow bands meet, that is, for $F_a$ around $5\times 10^{11}$ GeV
and $m_\sigma$ in the range $10^{6-7}$ GeV.

Before closing this section, let us comment on the lightest supersymmetric particle (LSP)
which is stable under the R-parity conservation.
In high scale SUSY, the thermal relic abundance of the LSPs may be too large.
This can be avoided if $T_\sigma$ is lower than the LSP freeze-out temperature.
If LSPs are non-thermally produced by saxion decays, its branching fraction must be suppressed.
Another way is to consider some mediation mechanism rendering the LSP sufficiently light,
or to introduce a very light superparticle in a hidden sector into which the ordinary
superparticles decay.
Alternatively one can simply assume R-parity violation.
We also note that the results on the baryon asymmetry and dark matter in the saxion-dominated
universe hold even in non-supersymmetric axion models,
if the saxion is stabilized with $m_\sigma\ll f$.

\section{Saxion stabilization}
\label{sec:saxion-stabilization}

In this section we discuss how to make the saxion dominantly decay into
right-handed neutrinos rather than into axions.
This requires a small $r$ in (\ref{Br-Ni}).
If PQ symmetry is spontaneously broken by a single field $X$,
which is the case for instance in the radiative saxion stabilization,
one obtains $r=1$, implying that ${\rm Br}(\sigma\to N_iN_i)$
is much smaller than $B_a$.
Even in this case, one can avoid an axion-dominated universe
by introducing the superpotential term, $X^2 H_uH_d$, so that
the saxion dominantly decays into SM particles or higgsinos.
However small ${\rm Br}(\sigma\to N_iN_i)$ makes it difficult to
implement non-thermal leptogenesis successfully.
This is resolved when multiple fields are involved
in the PQ symmetry breakdown.

Let us consider supergravity models where two PQ fields, $X_1$ and $X_2$,
obtain VEVs around $f$.
One simple way to stabilize the saxion is to consider
\bea
W = y_X \Sigma (X_1X_2 -\mu^2),
\eea
with $q_1+q_2=0$ and $\mu\sim f$, where $\Sigma$ is a PQ singlet field.
In this case, the $F$-flat direction $X_1X_2=\mu^2$ is lifted by
soft SUSY breaking terms,
\bea
-{\cal L}_{\rm soft} = m^2_1 |X_1|^2+m^2_2|X_2|^2.
\eea
For $y_X\mu \gg m_{3/2}$ and positive $m^2_{1,2}$ similar to or smaller than $m^2_{3/2}$,
the PQ fields are stabilized at
\bea
m^2_1 v^2_1 \simeq m^2_2 v^2_2,
\eea
with $v_1v_2=\mu^2$.
Thus, if $m^2_1=m^2_2$, the PQ fields are fixed with $v_1\simeq v_2$,
for which $r$ has a tiny value.
This implies that $r$ can have a small value when $X_1$ and $X_2$
feel SUSY breaking via couplings of similar strength.
For $X_{1,2}$ having soft scalar masses $m^2\pm \Delta m^2$
with $\Delta m^2\ll m^2$, we get $r \simeq \Delta m^2/m^2$.
If $\Delta m^2/m^2=0.1$, which would not require severe fine-tuning of
model parameters,
the relation (\ref{Br-Ni}) tells ${\rm Br}(\sigma\to N_iN_i)<3.6 B_a$.
In this case, one can arrange for instance ${\rm Br}(\sigma\to N_iN_i)$ around
0.7 and $B_a\simeq 0.2$ by taking appropriate right-handed neutrino masses.
Note that $B_N+B_a$ is not necessarily equal to one in the presence of
a superpotential term, $X^2_\alpha H_uH_d$.

Another way to stabilize the saxion is through the competition between
SUSY breaking effects and a higher dimensional superpotential term:
\bea
W = k \frac{X^{n_1}_1X^{n_2}_2}{\Lambda^{n_1+n_2-3}},
\eea
with $n_1+n_2\geq 4$ and $q_1n_1+q_2n_2=0$.
Here we take $q_1+q_2\neq 0$ to forbid a large mass term in the superpotential.
It is straightforward to find
\bea
q_1 v^2_1 = -q_2 v^2_2 \Big(1 + {\cal O}\Big(\frac{m^2_{1,2}}{m^2_{3/2}}\Big) \Big)
\sim
\left(\Lambda^{n_1+n_2-3} m_{3/2} \right)^{2/(n_1+n_2-2)},
\eea
and $m_\sigma\sim m_{3/2}$, assuming $|m^2_{1,2}|\ll m^2_{3/2}$.
The value of $r$ is fixed to be $r\simeq (1-n_1/n_2)^2$ or $(1-n_2/n_1)^2$,
depending on which PQ field is responsible for $M_i$.
Let us suppose that $N_i$ obtain masses from the coupling to one of the PQ fields
that gives a smaller value of $r$.
Then the saxion stabilization by the higher dimensional operator leads to
\bea
\label{saxion-models}
(n_1,n_2)=(1,3) \,\,&:&\,\,
r \simeq 0.44, \quad
f \sim \Lambda^{1/2} m^{1/2}_{3/2},
\nonumber \\
(n_1,n_2)=(2,3) \,\,&:&\,\,
r \simeq 0.11, \quad
f \sim \Lambda^{2/3} m^{1/3}_{3/2}.
\eea
The first model has ${\rm Br}(\sigma\to N_iN_i)<0.8B_a$ and $B_N<2.4B_a$.
Thus, to get a sufficient baryon asymmetry, one would need heavy $N_1$ with mass close
to $10^7$ GeV if the right-handed neutrino masses are hierarchical.
One may instead consider degenerate $N_i$ with $M\sim 10^6$ GeV and $M/\Delta M\sim 10$,
for which the successful non-thermal leptogenesis is achievable with
$B_a\sim 0.3$ and $B_N\sim 0.7$.
On the other hand, in the second model, we have ${\rm Br}(\sigma\to N_iN_i)<3.6B_a$ and
$B_N<10.8B_a$.
It is thus easy to make the saxion dominantly decay into right-handed neutrinos.
We also note that, in both cases, the PQ breaking scale $f$ naturally lies
in the range $10^{11-13}$ GeV for $m_{3/2}\sim 10^{6-7}$ GeV, taking $\Lambda$ to
be the GUT or Planck scale.\footnote{
The model discussed in Ref. \cite{Feldstein:2012bu} also realizes an intermediate axion
decay constant by relating it to the SUSY breaking scale, $F_a\sim \sqrt{m_{3/2}M_{Pl}}$.
However it is difficult to implement axionic co-genesis in this model
because the PQ sector itself participates in SUSY breaking, leading to a too large
saxion mass around $F_a$.
}

Our discussion so far has assumed that the saxion given by (\ref{saxion-field})
does not mix with other CP-even scalar bosons, and that its decay rate into the PQ sector
scalars or fermions is small compared to that into right-handed neutrinos.
For the model where $X_1$ and $X_2$ are stabilized by a higher dimensional superpotential
term, there appear additional light scalars and singlino.
Hence more care should be taken.
The saxion $\sigma$ indeed corresponds to the eigenstate in the limit of vanishing
soft scalar masses, and there is induced small mixing with the other CP-even scalar
for $|m^2_{1,2}|\ll m^2_{3/2}$.
Assuming small soft scalar masses, we find that the additional CP even and odd scalars
are both heavier than the saxion in the models (\ref{saxion-models}).
The PQ sector also includes two fermions composed of $\tilde X_1$ and $\tilde X_2$.
They have masses about $2.12m_\sigma$ and $0.71m_\sigma$ for $(n_1,n_2)=(1,3)$,
while $2.83m_\sigma$ and $0.71m_\sigma$ for the model with  $(n_1,n_2)=(2,3)$.
Thus the saxion decay into PQ fermions is kinematically closed in both models.

\section{Discussion and conclusions}

The PQ symmetry may be unbroken during inflation. Then the spontaneous breakdown of the
PQ symmetry after inflation leads to the production of topological defects such as
axionic strings and domain walls. According to the recent numerical simulations
of axion string-wall network~\cite{Hiramatsu:2010yu,Hiramatsu:2010yn}, the PQ breaking scale
should be of order $\GEV{10}$ for explaining the dark matter.
Based on our above argument, it will be difficult to combine this scenario with non-thermal
leptogenesis by the saxion decay.
See Fig.~\ref{fig:cogenesis}.
However, thermal leptogenesis will be possible, if the reheating temperature is sufficiently
high $T_R \gtrsim F_a \simeq \oten{10}$\,GeV.\footnote{
The thermal leptogenesis in the axion(=majoron) model was considered in Ref.~\cite{Langacker:1986rj}.
See also Ref.~\cite{Gu:2009hn}.
}
The axion will be thermalized, but its contribution to the effective number
of neutrinos is suppressed because of large relativistic degrees of freedom at high temperature.
Note that the saxion does not have to dominate the universe in this case, and the application
to a non-SUSY case is straightforward. It is interesting to note that the baryon and
dark matter abundances are determined by a single scale $M_1 \sim F_a \sim \oten{10}$\,GeV.

If the PQ symmetry is broken during inflation, the axion acquires quantum fluctuations, generating
isocurvature perturbations.
If the PQ breaking scale $F_a$ does not change during and after inflation, the inflation scale is
constrained by the isocurvature constraint as
$H_{\rm inf} \lesssim \oten{7}$\,GeV. This bound is relaxed if one of the PQ scalars takes a larger
field value during the inflation.
Such large deviation will ease the conditions for the saxion domination.
The isocurvature perturbations of the axion dark matter may be found or constrained by the future
observations of the CMB power spectrum and non-Gaussianity~\cite{Kawasaki:2008sn,Langlois:2008vk,
Kawakami:2009iu,Langlois:2011zz,Langlois:2010fe,Kobayashi:2013nva}.

In this paper we have considered the PQ sector as the origin of baryon asymmetry, dark matter and
dark radiation; {\it the axionic co-genesis}.
We have found that  non-thermal leptogenesis by the saxion decay works successfully
for the saxion mass of $\oten{6-7}$\,GeV and the PQ breaking scale around $5\times\GEV{11}$.
The dark matter can be explained by the axions produced in the misalignment mechanism with an initial
deviation of order unity.
If the existence of dark radiation is confirmed by future observations such as the Planck
satellite, this scenario will be one of the plausible solutions to the cosmological coincidence problems.
Intriguingly, in the SUSY framework, the suggested SUSY breaking scale is consistent with the observed
SM-like Higgs boson of mass $126$\,GeV in the minimal supersymmetric SM without large stop mixing.
Thus, the axionic co-genesis may give an answer to the question of why the SUSY breaking scale is much
higher than the weak scale, which generically requires severe fine-tuning for the correct electroweak
symmetry breaking.

\section*{Acknowledgment}
We thank Tetsutaro Higaki for pointing out a possibility to generate the right-handed neutrino
mass from the Giudice-Masiero mechanism.
This work was supported by the Grant-in-Aid for Scientific Research on Innovative
Areas (No.24111702, No. 21111006, and No.23104008) [FT], Scientific Research (A)
(No. 22244030 and No.21244033) [FT], and JSPS Grant-in-Aid for Young Scientists (B)
(No. 24740135) [FT]. This work was also supported by World Premier International Center Initiative
(WPI Program), MEXT, Japan [FT],
and by Grants-in-Aid for Scientific Research from the Ministry of Education, Science, Sports,
and Culture (MEXT), Japan, No. 23104008 and No. 23540283 [KJS].


\begin{thebibliography}{99}

\bibitem{Kaplan:2009ag}
  D.~E.~Kaplan, M.~A.~Luty and K.~M.~Zurek,
  Phys.\ Rev.\ D {\bf 79}, 115016 (2009)
  [arXiv:0901.4117 [hep-ph]].


\bibitem{Peccei:1977hh}
  R.~D.~Peccei and H.~R.~Quinn,
  Phys.\ Rev.\ Lett.\  {\bf 38}, 1440 (1977);
  Phys.\ Rev.\ D {\bf 16}, 1791 (1977).

\bibitem{QCD-axion}
  For a review, see
  J.~E.~Kim,
  Phys.\ Rept.\  {\bf 150}, 1 (1987);
  H.~Y.~Cheng,
  Phys.\ Rept.\  {\bf 158}, 1 (1988);
  J.~E.~Kim and G.~Carosi,
  Rev.\ Mod.\ Phys.\ \ {\bf 82}, 557  (2010)
  [arXiv:0807.3125 [hep-ph]];
 A.~Ringwald,
 Phys.\ Dark Univ.\  {\bf 1} (2012) 116
 [arXiv:1210.5081 [hep-ph]];
  M.~Kawasaki and K.~Nakayama,
  arXiv:1301.1123 [hep-ph]. 


\bibitem{Fukugita:1986hr}
  M.~Fukugita, T.~Yanagida,
  Phys.\ Lett.\  {\bf B174}, 45 (1986).

\bibitem{Buchmuller:2005eh}
  W.~Buchmuller, R.~D.~Peccei and T.~Yanagida,
  Ann.\ Rev.\ Nucl.\ Part.\ Sci.\  {\bf 55}, 311 (2005)  [hep-ph/0502169].  

\bibitem{Langacker:1986rj}
  P.~Langacker, R.~D.~Peccei and T.~Yanagida,
  Mod.\ Phys.\ Lett.\ A {\bf 1}, 541 (1986).

\bibitem{Lazarides:1991wu}
  G.~Lazarides and Q.~Shafi,
  Phys.\ Lett.\ B {\bf 258}, 305 (1991).

\bibitem{Asaka:1999yd}
  T.~Asaka, K.~Hamaguchi, M.~Kawasaki, T.~Yanagida,
  Phys.\ Lett.\  {\bf B464}, 12-18 (1999)
  [hep-ph/9906366];
  Phys.\ Rev.\  {\bf D61}, 083512 (2000)
  [hep-ph/9907559].


\bibitem{Chun:2000jr}
  E.~J.~Chun, D.~Comelli and D.~H.~Lyth,
  Phys.\ Rev.\ D {\bf 62}, 095013 (2000)
  [hep-ph/0008133].

\bibitem{Ichikawa:2007jv}
K.~Ichikawa, M.~Kawasaki, K.~Nakayama, M.~Senami and F.~Takahashi,
JCAP {\bf 0705} (2007) 008
[arXiv:hep-ph/0703034].



\bibitem{Higaki:2012ba}
  T.~Higaki, K.~Kamada and F.~Takahashi,
  JHEP {\bf 1209}, 043 (2012)
  [arXiv:1207.2771 [hep-ph]].

\bibitem{Higaki:2012ar}
  T.~Higaki and F.~Takahashi,
  JHEP {\bf 1211}, 125 (2012)  [arXiv:1208.3563 [hep-ph]].  

\bibitem{Cicoli:2012aq}
  M.~Cicoli, J.~P.~Conlon and F.~Quevedo,
  arXiv:1208.3562 [hep-ph].  

\bibitem{Fischler:2010xz}
  W.~Fischler and J.~Meyers,
  Phys.\ Rev.\ D {\bf 83}, 063520 (2011)
  [arXiv:1011.3501 [astro-ph.CO]].

\bibitem{Hasenkamp:2011em}
  J.~Hasenkamp,
  Phys.\ Lett.\ B {\bf 707}, 121 (2012)
   [arXiv:1107.4319 [hep-ph]].

\bibitem{Menestrina:2011mz}
  J.~L.~Menestrina and R.~J.~Scherrer,
  Phys.\ Rev.\ D {\bf 85}, 047301 (2012)
  [arXiv:1111.0605 [astro-ph.CO]].

\bibitem{Kobayashi:2011hp}
  T.~Kobayashi, F.~Takahashi, T.~Takahashi and M.~Yamaguchi,
  JCAP {\bf 1203}, 036 (2012)
  [arXiv:1111.1336 [astro-ph.CO]].

\bibitem{Hooper:2011aj}
  D.~Hooper, F.~S.~Queiroz and N.~Y.~Gnedin,
  Phys.\ Rev.\ D {\bf 85}, 063513 (2012)
  [arXiv:1111.6599 [astro-ph.CO]].

\bibitem{Jeong:2012np}
  K.~S.~Jeong and F.~Takahashi,
  JHEP {\bf 1208}, 017 (2012)  [arXiv:1201.4816 [hep-ph]].  


\bibitem{Choi:2012zna}
  K.~Choi, K.~-Y.~Choi and C.~S.~Shin,
  Phys.\ Rev.\ D {\bf 86}, 083529 (2012)  [arXiv:1208.2496 [hep-ph]].  


\bibitem{Graf:2012hb}
  P.~Graf and F.~D.~Steffen,
  arXiv:1208.2951 [hep-ph].  


\bibitem{Hasenkamp:2012ii}
  J.~Hasenkamp and J.~Kersten,
  arXiv:1212.4160 [hep-ph].  


\bibitem{Bae:2013qr}
  K.~J.~Bae, H.~Baer and A.~Lessa,
  arXiv:1301.7428 [hep-ph].  


\bibitem{Nakayama:2010vs}
K.~Nakayama, F.~Takahashi and T.~T.~Yanagida,
Phys.\ Lett.\ B {\bf 697} (2011) 275
[arXiv:1010.5693 [hep-ph]].


\bibitem{Hou:2012xq}
  Z.~Hou, C.~L.~Reichardt, K.~T.~Story, B.~Follin, R.~Keisler, K.~A.~Aird, B.~A.~Benson and L.~E.~Bleem {\it et al.},
  arXiv:1212.6267 [astro-ph.CO].  

\bibitem{Hinshaw:2012fq}
  G.~Hinshaw, D.~Larson, E.~Komatsu, D.~N.~Spergel, C.~L.~Bennett, J.~Dunkley, M.~R.~Nolta and M.~Halpern {\it et al.},
  arXiv:1212.5226 [astro-ph.CO].  

\bibitem{Ade:2013lta}
  P.~A.~R.~Ade {\it et al.}  [Planck Collaboration],
  arXiv:1303.5076 [astro-ph.CO].


\bibitem{:2012gk}
  G.~Aad {\it et al.}  [ATLAS Collaboration],
  Phys.\ Lett.\ B {\bf 716}, 1 (2012)
  [arXiv:1207.7214 [hep-ex]].  

\bibitem{:2012gu}
  S.~Chatrchyan {\it et al.}  [CMS Collaboration],
  Phys.\ Lett.\ B {\bf 716}, 30 (2012)
  [arXiv:1207.7235 [hep-ex]].  


\bibitem{Okada:1990gg}
  Y.~Okada, M.~Yamaguchi and T.~Yanagida,
  Phys.\ Lett.\ B {\bf 262}, 54 (1991);
  see also
  Y.~Okada, M.~Yamaguchi and T.~Yanagida,
  Prog.\ Theor.\ Phys.\  {\bf 85}, 1 (1991);
  J.~R.~Ellis, G.~Ridolfi and F.~Zwirner,
  Phys.\ Lett.\ B {\bf 257}, 83 (1991);
  H.~E.~Haber and R.~Hempfling,
  Phys.\ Rev.\ Lett.\  {\bf 66}, 1815 (1991).

\bibitem{Giudice:2011cg}
  G.~F.~Giudice and A.~Strumia,
  Nucl.\ Phys.\ B {\bf 858}, 63 (2012)
  [arXiv:1108.6077 [hep-ph]];\\
  G.~Degrassi, S.~Di Vita, J.~Elias-Miro, J.~R.~Espinosa, G.~F.~Giudice, G.~Isidori and A.~Strumia,
  JHEP {\bf 1208}, 098 (2012)
  [arXiv:1205.6497 [hep-ph]];\\
see also
  F.~Bezrukov, M.~Y.~.Kalmykov, B.~A.~Kniehl and M.~Shaposhnikov,
  JHEP {\bf 1210}, 140 (2012)
  [arXiv:1205.2893 [hep-ph]].


\bibitem{seesaw}
T.~Yanagida, in Proceedings of the {\it{``Workshop on the Unified Theory and
 the Baryon Number in the universe''}}, Tsukuba, Japan, Feb. 13-14, 1979, edited by
O.~Sawada and A.~Sugamoto, KEK report KEK-79-18, p. 95,
and {\it{``Horizontal Symmetry And Masses Of Neutrinos''
}}, Prog. Theor. Phys. {\bf{64}} (1980) 1103;
M.~Gell-Mann, P.~Ramond and R.~Slansky, in {\it{``Supergravity''}}
 (North-Holland, Amsterdam, 1979) {\it{eds}}. D.~Z.~Freedom and P.~van
Nieuwenhuizen, Print-80-0576 (CERN);
see also   P.~Minkowski,  Phys.\ Lett.\  B {\bf 67}, 421 (1977).


\bibitem{Chun:1995hc}
  E.~J.~Chun and A.~Lukas,
  Phys.\ Lett.\ B {\bf 357}, 43 (1995)
  [hep-ph/9503233].

\bibitem{Giudice:1988yz}
  G.~F.~Giudice and A.~Masiero,
  Phys.\ Lett.\ B {\bf 206}, 480 (1988).  

\bibitem{Kim:1983dt}
  J.~E.~Kim and H.~P.~Nilles,
  Phys.\ Lett.\  B {\bf 138}, 150 (1984).


\bibitem{Kawasaki:2011ym}
  M.~Kawasaki, N.~Kitajima and K.~Nakayama,
  Phys.\ Rev.\ D {\bf 83}, 123521 (2011)
  [arXiv:1104.1262 [hep-ph]].


\bibitem{Feldstein:2012bu}
  B.~Feldstein and T.~T.~Yanagida,
  arXiv:1210.7578 [hep-ph].  


\bibitem{Hiramatsu:2010yu}
  T.~Hiramatsu, M.~Kawasaki, T.~Sekiguchi, M.~Yamaguchi and J.~'i.~Yokoyama,
  Phys.\ Rev.\ D {\bf 83}, 123531 (2011)
  [arXiv:1012.5502 [hep-ph]].

\bibitem{Hiramatsu:2010yn}
  T.~Hiramatsu, M.~Kawasaki and K.~'i.~Saikawa,
  JCAP {\bf 1108}, 030 (2011)
  [arXiv:1012.4558 [astro-ph.CO]].


\bibitem{Graf:2010tv}
  P.~Graf and F.~D.~Steffen,
  Phys.\ Rev.\ D {\bf 83}, 075011 (2011)
  [arXiv:1008.4528 [hep-ph]].


\bibitem{Choi:1996vz}
  K.~Choi, E.~J.~Chun and J.~E.~Kim,
  Phys.\ Lett.\  B {\bf 403}, 209 (1997)
  [arXiv:hep-ph/9608222].


\bibitem{Turner:1985si}
  M.~S.~Turner,
  Phys.\ Rev.\ D {\bf 33}, 889 (1986);  
  K.~J.~Bae, J.~-H.~Huh and J.~E.~Kim,
  JCAP {\bf 0809}, 005 (2008) [arXiv:0806.0497 [hep-ph]].


\bibitem{Gross:1980br}
  D.~J.~Gross, R.~D.~Pisarski and L.~G.~Yaffe,
  Rev.\ Mod.\ Phys.\  {\bf 53}, 43 (1981).


\bibitem{Covi:1996wh}
  M.~Flanz, E.~A.~Paschos and U.~Sarkar,
  Phys.\ Lett.\ B {\bf 345}, 248 (1995)  [hep-ph/9411366];  
%
  L.~Covi, E.~Roulet and F.~Vissani,
  Phys.\ Lett.\ B {\bf 384}, 169 (1996)  [hep-ph/9605319];  
  W.~Buchmuller and M.~Plumacher,
  Phys.\ Lett.\ B {\bf 431}, 354 (1998)  [hep-ph/9710460].  



\bibitem{Gu:2009hn}
  P.~-H.~Gu and U.~Sarkar,
  Eur.\ Phys.\ J.\ C {\bf 71}, 1560 (2011)
  [arXiv:0909.5468 [hep-ph]].


\bibitem{Kawasaki:2008sn}
  M.~Kawasaki, K.~Nakayama, T.~Sekiguchi, T.~Suyama and F.~Takahashi,
  JCAP {\bf 0811}, 019 (2008)
  [arXiv:0808.0009 [astro-ph]];
  JCAP {\bf 0901}, 042 (2009)
  [arXiv:0810.0208 [astro-ph]].

\bibitem{Langlois:2008vk}
  D.~Langlois, F.~Vernizzi and D.~Wands,
  JCAP {\bf 0812}, 004 (2008)
  [arXiv:0809.4646 [astro-ph]].

\bibitem{Kawakami:2009iu}
  E.~Kawakami, M.~Kawasaki, K.~Nakayama, F.~Takahashi and ,
  JCAP {\bf 0909}, 002 (2009)
  [arXiv:0905.1552 [astro-ph.CO]].

\bibitem{Langlois:2011zz}
  D.~Langlois, A.~Lepidi and ,
  JCAP {\bf 1101}, 008 (2011)
  [arXiv:1007.5498 [astro-ph.CO]].

\bibitem{Langlois:2010fe}
  D.~Langlois, T.~Takahashi and ,
  JCAP {\bf 1102}, 020 (2011)
  [arXiv:1012.4885 [astro-ph.CO]].

\bibitem{Kobayashi:2013nva}
  T.~Kobayashi, R.~Kurematsu and F.~Takahashi,
  arXiv:1304.0922 [hep-ph].

\end{thebibliography}
\end{document}